\documentclass[nofootinbib,superscriptaddress,a4paper, twocolumn,floatfix]{revtex4-2}
\usepackage[english]{babel}
\usepackage[utf8]{inputenc}
\usepackage[colorinlistoftodos, color=green!40, prependcaption]{todonotes}
\usepackage{amsthm}
\usepackage{mathtools}
\usepackage{physics}
\usepackage{braket}
\usepackage{listings}
\usepackage{xcolor}
\usepackage{graphicx}
\usepackage[left=23mm,right=13mm,top=35mm,columnsep=15pt]{geometry} 
\usepackage{adjustbox}
\usepackage{placeins}
\usepackage[T1]{fontenc}
\usepackage{lipsum}
\usepackage{csquotes}
\usepackage[pdftex, pdftitle={Article}, pdfauthor={Author}]{hyperref} 
\usepackage{booktabs}
\usepackage{amssymb}
\usepackage{bm}
\usepackage[ruled,vlined]{algorithm2e}
\usepackage[caption=false]{subfig}
\usepackage{duckuments}
\usepackage{qcircuit}

\usepackage{float}

\makeatletter
\newcommand*{\rom}[1]{\expandafter\@slowromancap\romannumeral #1@}
\makeatother

\begin{document}


\title{\textbf{Explainable Representation Learning of Small Quantum States}}


\author{Felix Frohnert}
\email[E-mail:]{f.frohnert@liacs.leidenuniv.nl}
\affiliation{$\langle a Q a^L \rangle$ Applied Quantum Algorithms, Universiteit Leiden}

\author{Evert van Nieuwenburg}
\affiliation{$\langle a Q a^L \rangle$ Applied Quantum Algorithms, Universiteit Leiden}

\date{\today}

\begin{abstract}
Unsupervised machine learning models build an internal representation of their training data without the need for explicit human guidance or feature engineering.
This learned representation provides insights into which features of the data are relevant for the task at hand.
In the context of quantum physics, training models to describe quantum states without human intervention offers a promising approach to gaining insight into how machines represent complex quantum states.  
The ability to interpret the learned representation may offer a new perspective on non-trivial features of quantum systems and their efficient representation. 
We train a generative model on two-qubit density matrices generated by a parameterized quantum circuit.
In a series of computational experiments, we investigate the learned representation of the model and its internal understanding of the data. 
We observe that the model learns an interpretable representation which relates the quantum states to their underlying entanglement characteristics.
In particular, our results demonstrate that the latent representation of the model is directly correlated with the entanglement measure concurrence. 
The insights from this study represent proof of concept towards interpretable machine learning of quantum states.  
Our approach offers insight into how machines learn to represent small-scale quantum systems autonomously. 
\end{abstract}

\maketitle

\section{Introduction}
Over the past decades, (un)supervised representation learning has revolutionized machine learning research~\cite{bengioRepresentationLearningReview2013a}. 
While manual feature engineering with specific domain expertise used to be required~\cite{hofmannKernelMethodsMachine2008}, powerful deep neural networks have proven to be successful in automatically extracting useful representations of data.
This advent has led to better performance on a wide range of tasks, such as language modeling and computer vision~\cite{karrasProgressiveGrowingGANs2018,devlinBERTPretrainingDeep2019,bowmanGeneratingSentencesContinuous2016a}.
In recent years, the application of representation learning has found its way into the physical sciences. 
It has been applied to studying phases of matter \cite{wetzelUnsupervisedLearningPhase2017,kottmannUnsupervisedPhaseDiscovery2020, tibaldiUnsupervisedSupervisedLearning2023}, detection of outliers in particle collision experiments \cite{farinaSearchingNewPhysics2020,cerriVariationalAutoencodersNew2019}, learning spectral functions \cite{milesMachineLearningKondo2021}, and compression of quantum states~\cite{rocchettoLearningHardQuantum2018}.
The last category, in particular, raises the interesting question of which properties of quantum systems are deemed important to capture by the machine learning model when compressing them. 
By examining and interpreting salient features of the learned representation built without human intervention, we can uncover the models internal understanding of a quantum physical system.    
Adding the constraint of learning representations which are meaningful ~\cite{nautrupOperationallyMeaningfulRepresentations2022} and explainable ~\cite{barredoarrietaExplainableArtificialIntelligence2020} is an important prerequisite for the development of an artificial intelligence system for physics research.
The incorporation of this constraint serves as a vital prerequisite for achieving the ultimate goal of building artificial intelligence systems that can facilitate new scientific discoveries~\cite{itenDiscoveringPhysicalConcepts2020a,nautrupOperationallyMeaningfulRepresentations2022}.

In this work, we focus on studying two-qubit quantum circuits in the presence of information scrambling and depolarization, and investigate if a generative model~\cite{goodfellowGenerativeAdversarialNetworks2014} is able to learn representations highlighting entanglement features. 
We apply local information scrambling to the states to inhibit the model's ability to exploit local features for the purpose of identifying the generative parameter, while simultaneously preserving the non-local entanglement properties.
We therefore follow the recent development of training generative models to discover interpretable physical representations~\cite{milesMachineLearningKondo2021,flam-shepherdLearningInterpretableRepresentations2022a,routhLatentRepresentationLearning2021,luExtractingInterpretablePhysical2020,kalininExploringOrderParameters2021,liuRepresentationLearningQuantum2022,zhuFlexibleLearningQuantum2022} and want to investigate the following question: Given a set of quantum states that differ solely in their entanglement properties, how will an unsupervised machine learning model learn to represent them?

We encode the full density matrices generated by two-qubit circuits using a variational autoencoder (VAE), which has been established as a suitable model for learning meaningful internal representations~\cite{kingmaAutoEncodingVariationalBayes2022a}.
This is schematically depicted in Fig.~\ref{fig:vae_overview}.
A VAE performs dimensionality reduction~\cite{vandermaaten2009dimensionality}, compressing an input into a smaller dimension called the latent space, and then attempts to reconstruct the input from that latent representation.
Originally proposed as a generative model for image data, this architecture has proven to be capable of extracting ground-truth generative factors (underlying feature of the data which captures a distinct attribute or characteristic) from highly complex feature spaces and representing them in a human interpretable manner~\cite{salakhutdinovLearningDeepGenerative2015a,higginsBetaVAELearningBasic2022a}. 
In particular, the so-called $\beta$-VAE introduces a regularization hyperparameter which encourages independent latent variables, leading to more interpretable representations~\cite{higginsBetaVAELearningBasic2022a}.
Thus, we conduct a hyperparameter search on $\beta$, and our results reveal that the smallest latent representation the model can learn is interpretable and captures entanglement properties. 
Specifically, our investigation shows that the latent space encodes a quantity which effectively follows known entanglement measures such as concurrence and negativity, which are identical for the two-qubit systems we focus on.
Moreover, we show that the model generalizes to any other two-qubit state, as well as to two-qubit subsets of three-qubit states.


The remainder of this paper is structured as follows.
In section~\ref{sec:circuit}, we present a description of the two-qubit system under consideration, including information about the corresponding data sets that were generated.
Additionally, we give a brief introduction to variational autoencoders.
In section~\ref{sec:result}, we present the results of experiments on density matrices with and without information scrambling.
We furthermore test the ability of the model to generalize to different quantum states.
We provide a thorough analysis of the learned representations and explore their relationship to the underlying properties of quantum states.
Finally, in section~\ref{sec:discussion}, we conclude the results and provide an overview on future work that can be undertaken to extend and improve upon the results presented in this paper.

\begin{figure}[t]
    \centering
    \includegraphics[width=1\linewidth]{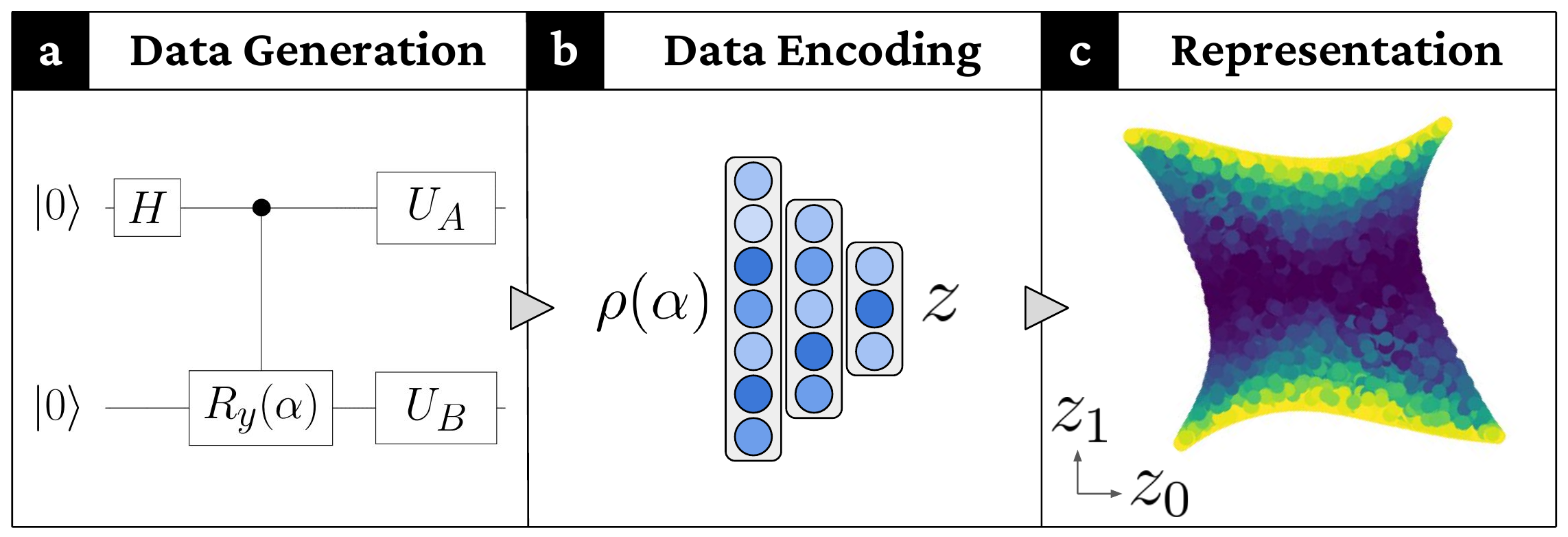}
\caption{ \textbf{Conceptual overview.} 
a) Quantum states $\rho(\alpha)$ are generated by a two-qubit quantum circuit consisting of a Hadamard, a Controlled-$R_y$ gate parameterized by the angle $\alpha$, and two single-qubit rotations. 
b) Data are encoded from a density matrix into a stochastic latent representation $z$ using the trained encoder network.
c) Latent variables $z=(z_0,z_1)$ are visualized to analyze the relation of structure of the learned representation and encoded properties. 
In this figure, the two-dimensional latent space is color-coded by an entanglement measure of underlying states (the concurrence). Here, the low entanglement region is colored purple and the high entanglement region is colored yellow.}
\label{fig:overview}
\end{figure}

\section{Methods} \label{sec:circuit}
\subsection{Data}\label{chap:data}
We study quantum states generated by the two-qubit parameterized quantum circuit in Fig. \ref{fig:overview}a ~\cite{cerezoVariationalQuantumAlgorithms2021}.  
This circuit consists of a Hadamard gate and a Controlled-$R_Y(\alpha)$ rotation with input angle $\alpha$, which produces the density matrix $\rho(\alpha)$ (see~\ref{app:calculation} for the full description). The random unitaries $U_A$ and $U_B$ will be discussed shortly.
For such a two-qubit system, the amount of entanglement can be quantified through the concurrence~\cite{nielsenQuantumComputationQuantum2010}:
\begin{equation}
    C[\rho(\alpha)] = \text{max}(0,\lambda_1-\lambda_2-\lambda_3-\lambda_4) \label{eq:conc}
\end{equation}
In this, $\lambda_i$ are the eigenvalues (in descending order) of the Hermitian matrix:
\begin{equation}
    R=\sqrt{\sqrt{\rho(\alpha)}   ( \sigma_y \otimes \sigma_y)\rho(\alpha)^*(\sigma_y \otimes \sigma_y)   \sqrt{\rho(\alpha)}}. \label{eq:r}
\end{equation}
At $\alpha = 0$ the state is fully separable (and hence $C[\rho(0)] = 0$), while for any non-zero $\alpha$ the state is entangled and has non-zero concurrence. This is visualized in Fig.~\ref{fig:conc} in the appendices. 
The motivation for choosing to study the states $\rho(\alpha)$ is that a single parameter $\alpha$ uniquely determines the entanglement properties of each state, and drawing $\alpha \in [0,\pi]$ explores the entire range of entanglement measure values.
This simple structure-property relation makes it easy to interpret learned representations.

\subsection{Variational Autoencoders}\label{chap:vae}
\begin{figure}[t]
    \centering
    \includegraphics[width=0.75\linewidth]{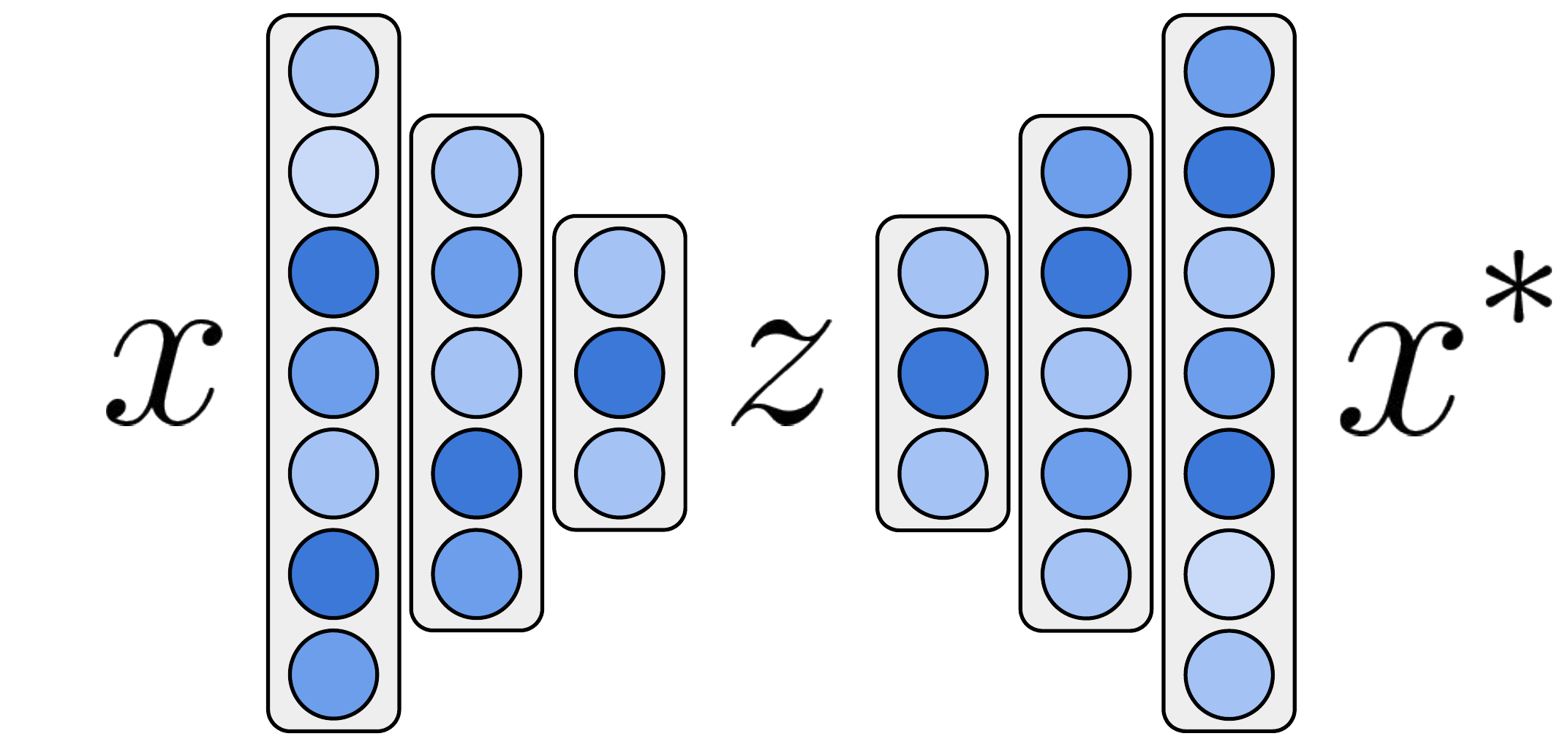}
\caption{\textbf{Schematic overview of VAE architecture.} The input $\textbf{x}$ is compressed by the neural network-based encoder into the latent space, represented as $\textbf{z}$, serving as an information bottleneck. The decoder network then uses the information from the latent space to reconstruct $\textbf{x}^*$.
}
\label{fig:vae_overview}
\end{figure}
Variational autoencoders aim to find an effective compressed representation of data by learning the identity map through an informational bottleneck~\cite{kingmaAutoEncodingVariationalBayes2022a}. 
As visualized in Fig. \ref{fig:vae_overview}, VAEs accomplish this task by using an encoder and decoder network. 
The encoder $q_{\bm{\phi}}(\bm{z}|\bm{x})$ is a neural network with weights and biases $\bm{\phi}$ that maps high-dimensional data to the so-called latent space:
\begin{equation}
    q_{\bm{\phi}}(\bm{z}|\bm{x}) = \mathcal{N}(\bm{z}|\bm{\mu }_ {\bm{\phi}}  (\bm{x}),  \bm{\sigma}^2_{\bm{\phi}} (\bm{x}) ). \label{eq:encoder}
\end{equation}
From a given data point $\bm{x}$ it generates a normal distribution $\mathcal{N}$ over the possible values of the latent variable $\bm{z} \sim q_{\bm{\phi}}(\bm{z}|\bm{x})$, from which $\bm{x}$ could have been generated.
In this, $\bm{z}=[z_0,\cdots,z_N]$ is a point in an $N$-dimensional latent space, where $N$ is chosen manually beforehand.
Though with an arbitrarily complex encoder a dataset can \emph{in principle} be encoded in just one latent variable\footnote{For example by performing a simple enumeration for a static dataset, or by a complex nonlinear encoding.}, \emph{in practice} a well-trained latent representation captures ground-truth generative factors in the input data~\cite{burgessUnderstandingDisentanglingBeta2018}. 
In our case, the encoder is a fully connected feedforward neural network consisting of multiple hidden layers with nonlinear activation functions
\begin{equation}
    \bm{\mu}_ {\bm{\phi}}, \bm{\sigma}^2_{\bm{\phi}} =  \text{MLP}_{\bm{\phi}} (\bm{x}).
\end{equation}
The mean $\bm{\mu }_ {\bm{\phi}}$ and variance $\bm{\sigma}^2_{\bm{\phi}}$ are the learned parameters defining the distribution in Eq. \ref{eq:encoder}.
For the visualization of the learned latent variables throughout the remainder of this manuscript, we will exclude their variance and instead concentrate solely on the mean values of their latent distributions, denoted as $\bm{z}=\bm{\mu}$. Assigning an interpretation solely to the mean is a common practice, as exemplified in  ~\cite{zhouLearningIdentifiableInterpretable2020}.



Similarly, the decoder $p_{\bm{\theta}}(\bm{x}|\bm{z})$ is a neural network with weights and biases $\bm{\theta}$ that attempts to reconstruct the input $\bm{x}$ from given latent variables $\bm{z}$ and follows the reversed structure of the encoder as shown in Fig. \ref{fig:vae_overview}.

During training the parameters $\bm{\phi}$ and $\bm{\theta}$ are tuned with the goal of minimizing the following loss function:
\begin{equation}
    \mathcal{L}(\bm{x};\bm{\phi},\bm{\theta}) =  \mathcal{L}_{R}(\bm{x};\bm{\phi},\bm{\theta}) + \beta \cdot  \mathcal{L}_{KL}(\bm{x};\bm{\phi},\bm{\theta}) \label{bvae_loss}
\end{equation}
This loss function is composed of two terms: a reconstruction loss $\mathcal{L}_{R}$ and a regularization loss $\mathcal{L}_{KL}$. 
The reconstruction loss measures the difference between the original input and the output of the decoder.
In our case, the metric for this difference will be the element-wise mean squared error of the input density matrices. 
This choice of metric influences the results, because with this metric the off-diagonal elements of the density matrix have a larger relative contribution. 
The regularization loss, on the other hand, is given by the Kullback-Leibler divergence of the latent representation and a standard normal distribution. 
This encourages the latent representation to be smooth and continuous, and moreover aims at having latent variables represent independent generative factors~\cite{kingmaAutoEncodingVariationalBayes2022a,higginsBetaVAELearningBasic2022a}. 
For a single data point this loss can be expressed as
\begin{equation}
    \mathcal{L}_{KL}=\sum_i^N \frac{1}{2} (\mu_i^2 + \sigma_i^2 -1 -\log \sigma_i^2) = \sum_i^N \mathcal{L}^{(i)}_{KL}\label{eq:kl},
\end{equation}
where $i$ runs over the $N$ latent variables.
The hyperparameter $\beta$ in Eq. \ref{bvae_loss} controls the impact of regularization on the overall optimization objective, regulating the trade-off between the effective encoding capacity of the latent space and the statistical independence of individual latent variables in the learned representation~\cite{higginsBetaVAELearningBasic2022a}.

\section{Results and Discussion}\label{sec:result}
In the following, we perform a series of experiments to evaluate VAE models with varying training data, latent dimensions, and $\beta$ regularization strengths. 
More details about training and model implementation can be found in section~\ref{chap:training_details}. 
We obtain a number of results from the experiments, which we discuss in the following.
%
%
This section initially focuses on encoding pure state density matrices without regularization, demonstrating the successful extraction of their generative parameter $\alpha$ using the VAE.
Next, an information scrambling technique is introduced to prevent the direct extraction of $\alpha$, and the optimization of the regularization parameter is shown to produce an interpretable representation closely following concurrence.
Finally, the section explores the generalization abilities of the VAE by investigating its performance on mixed states and three-qubit W states.

\subsection{Encoding Quantum States \boldmath$\rho(\alpha)$} \label{chap:exp_1}
In this investigation, we study how a VAE learns to encode pure state density matrices $\rho(\alpha)$ and refer to this specific model as $\rho$-VAE.
Though the data has one generative factor, we wish to explicitly confirm that one latent variable indeed suffices for reconstruction. 
To empirically confirm this, we train VAEs with different latent space dimensions ($N =[1,\ldots,8]$) on quantum states $\rho(\alpha)$, with $\alpha = [0,\pi]$ in $10^3$ steps and record the final loss $\mathcal{L}$. 
For each $N$, we run $9$ experiments and average the results.
Throughout the training process, we maintain a regularization strength of $\beta=0$.
\begin{figure}[t]
    \centering
    \includegraphics[width=1\linewidth]{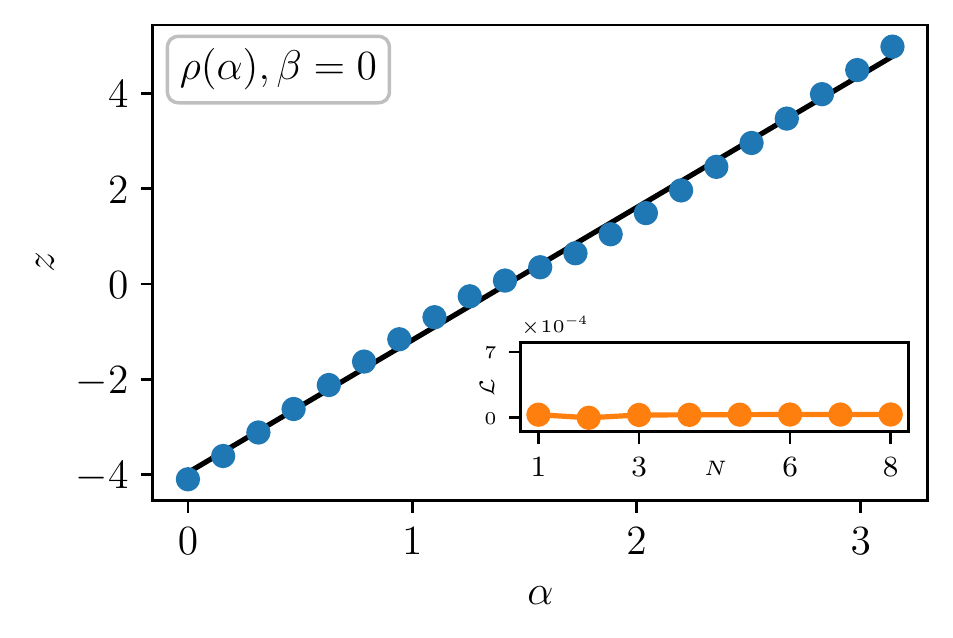}
    \caption{\textbf{The \boldmath$\rho$-VAE learns to extract the parameter \boldmath$\alpha$ from quantum states to structure its latent space.} \\
    The correlation between the one-dimensional latent space $z$ of $\rho$-VAE and parameter $\alpha$ of encoded density matrices (blue, mean and standard deviation of $10$ samples). The error bars are contained within the markers. 
    The regression of encoded quantum states (black) shows that the correlation has a small sinusoidal feature but is sufficiently characterized by a linear function with $r^2>0.99$.
    Inset: The final loss of $\rho$-VAE trained on quantum states $\rho(\alpha)$ at $\beta=0$ with latent space dimensions $N\in [1,8]$ (mean and standard deviation of $9$ experiments) indicates that a one-dimensional latent space has sufficient information capacity.
    }
    \label{fig:exp_1}
\end{figure}
The inset in Fig. \ref{fig:exp_1} shows these results, plotting the reconstruction quality of the trained model at different latent space sizes.
We find that indeed a one-dimensional (scalar) latent space is sufficient for compressing quantum states $\rho(\alpha)$, since increasing the number of latent variables does not lead to a significant decrease of the final loss. 

The next step of the analysis is to examine and interpret the learned representation of the one-dimensional model to uncover what property of quantum states it extracts to structure its latent space.
For this, we use the trained $\rho$-VAE ($N=1$) to encode a test set of quantum states at different $\alpha$ ($10$ samples at $21$ unique angles) and record the resulting $10$ predicted latent variables $z$.

Fig. \ref{fig:exp_1} shows the correlation between the mean of the predicted latent variable values (blue) and angle $\alpha$ of the corresponding input quantum states $\rho(\alpha)$. 
We find that the model assigns latent variable values that scale mostly linearly with the angle $\alpha$, as demonstrated by the linear regression with a coefficient of determination $r^2 > 0.99$~\cite{lewis2015applied}. 
In other words, the VAE extracts a latent parameter that is linearly correlated with the generative factor $\alpha$.
We note that there is no incentive for the VAE to extract the actual value of $\alpha$, as long as the latent representation can uniquely reconstruct inputs.

%
%

Finally, by investigating the structure of density matrices in Eq.~\ref{eq:dm}, we can also interpret why the model has learned to use this specific mapping from quantum state to latent representation: Each angle $\alpha \in [0,\pi]$ generates a density matrix with a unique structure, which means that extracting the generative angle $\alpha$ is a sufficient mapping of the sample to a single latent variable that allows for reconstruction.
%
%
As a final detail, we note that the predicted latent variable values exhibit a standard deviation near zero, with the error bars consistently falling within the markers. 
This observation shows the robustness of the model's predictions, indicating that an identical representation is consistently obtained across multiple experiments.

\subsection{Encoding Quantum States \boldmath$\rho_s(\alpha)$}\label{chap:exp_2}
In the next step, we introduce an information scrambling procedure to prevent the VAE from learning a direct map to the generative factor $\alpha$, and that additionally fully removes the ability to extract local features from quantum states.
In this experiment the density matrices are scrambled utilizing random local unitaries
\begin{equation}
     \rho_s(\alpha) = \left( U_A \otimes U_B\right) \rho(\alpha) \left( U_A \otimes U_B\right)^\dagger, \label{eq:scramble}
\end{equation}
where $U_A$ and $U_B$ are the real components of random local $2\times2$ unitary operators which are uniformly distributed according to the Haar measure~\cite{trávníček2023sensitivity,lundbergHaarMeasureGeneration2004a}. 
The procedure to generate these unitaries is detailed in Appendix~\ref{chap:rhos}. 
By applying the unitary transformation in Eq.~\ref{eq:scramble} to the density matrices, local information becomes inaccessible while non-local information remains invariant~\cite{gavreevLearningEntanglementBreakdown2022a}. 



\begin{figure}[t]
    \centering
    \includegraphics[width=1\linewidth]{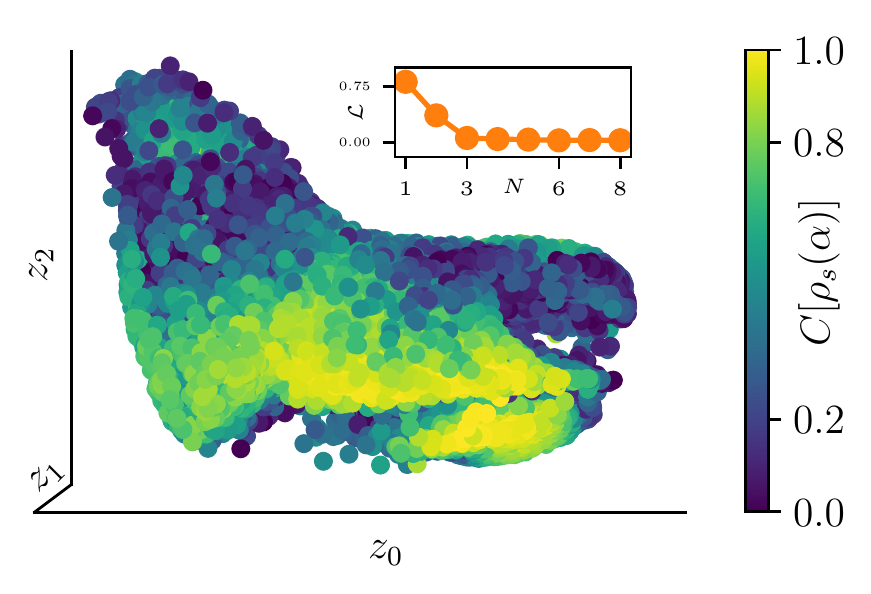}
    \caption{\textbf{The \boldmath$\rho_s$-VAE learns an efficient but uninterpretable representation of quantum states with information scrambling \boldmath$\rho_s(\alpha)$.}
    Three-dimensional latent space $z=(z_0,z_1,z_2)$ of $\rho_s$-VAE 
    trained with $\beta=0$. Each encoded density matrix is color-coded by its corresponding concurrence value.
    Inset: The final loss of $\rho_s$-VAE trained on quantum states $\rho_s(\alpha)$ at $\beta=0$ with latent space dimensions $N\in [1,8]$ (mean and standard deviation of $9$ experiments) indicates that a three-dimensional latent space has sufficient information capacity. 
    }
    \label{fig:exp_2}
\end{figure}

We now study how a VAE, which we label the $\rho_s$-VAE, learns to encode the scrambled density matrices $\rho_s(\alpha)$, keeping $\beta = 0$.
The inset in Fig.~\ref{fig:exp_2} again illustrates the change in reconstruction quality of the trained model at different latent space sizes.
A perfect reconstruction would require extracting $7$ generative factors: the angle $\alpha$ and $3$ angles each for the random unitaries.
And though the lowest loss values are indeed at $N \leq 7$, we observe a clear kink at a three-dimensional latent space, after which the loss flattens out.
%
We are not after a perfect reconstruction but rather focus on interpretable latent spaces, and hence a smaller latent space is preferred over exact reconstruction.

To examine and attempt to interpret the learned representation, we encode a test set of $\rho_s(\alpha)$ quantum states using the trained $N = 3$ model and record the predicted latent variable values $z$.

Fig.~\ref{fig:exp_2} visualizes the latent encoding of quantum states, where each point is color-coded by the concurrence value $C[\rho_s(\alpha)]$. 
We note that this representation is structured by regions of high entanglement (yellow), minimal entanglement (purple), and mixed regions. 
This observation suggests that the model constructs its latent space according to some underlying properties of the quantum states.
%
However, one caveat of this representation is that the extracted information is shared between the three latent dimensions, as all of them appear to capture certain aspects of non-local properties.
This makes it impossible to readily interpret the latent variables and to derive a general statement about the learned map from sample to latent representation, which is a well-known problem of VAEs with non-optimized regularization strength~\cite{higginsBetaVAELearningBasic2022a}.

\subsection{Tuning Regularization Strength \boldmath$\beta$ }\label{chap:exp_3}
Hence, to optimize for interpretability, we tune the regularization strength $\beta$ of the $\rho_s$-VAE.
The goal is to find a representation with factorized (disentangled) latent variables, meaning that each latent dimension represents a unique independent feature of the encoded data. 
This is beneficial for us, as representation in which the latent variables learn to encode different independent generative factors of variation in the data is better tuned to human intuition in interpreting data compared to the previous standard VAE approach~\cite{burgessUnderstandingDisentanglingBeta2018}.

By adjusting the value of $\beta$, we can control how much the latent variables resemble a normal distribution throughout the optimization process. 
This naturally incorporates the properties of the normal prior into the learned representation, such as its factorized nature.
Specifically, the characteristic of a diagonal covariance matrix of the latent variables is advantageous for the goal of finding interpretable representations, as it creates a disentangled latent space in which each dimension is independent and uncorrelated with the others.
Importantly, the tuning process leads to a trade-off between the reconstruction quality of the encoded input and the degree of disentanglement of the learned latent space, where a higher value of $\beta$ generally leads to more disentangled latent variables but lower reconstruction quality \cite{higginsBetaVAELearningBasic2022a}.

We train the $\rho_s$-VAE on quantum states $\rho_s(\alpha)$ with $\beta$ ranging from $0.01$ to $1.2$ with a large latent space of dimension $N=8$ to give the latent bottleneck sufficient capacity.
For each value of $\beta$, we train a model and record the regularization loss value $\mathcal{L}^{(i)}_{KL}$ (see Eq.~\ref{eq:kl}) of each latent variable averaged across the data set.
\begin{figure}[t]
    \centering
    \includegraphics[width=1.0\linewidth]{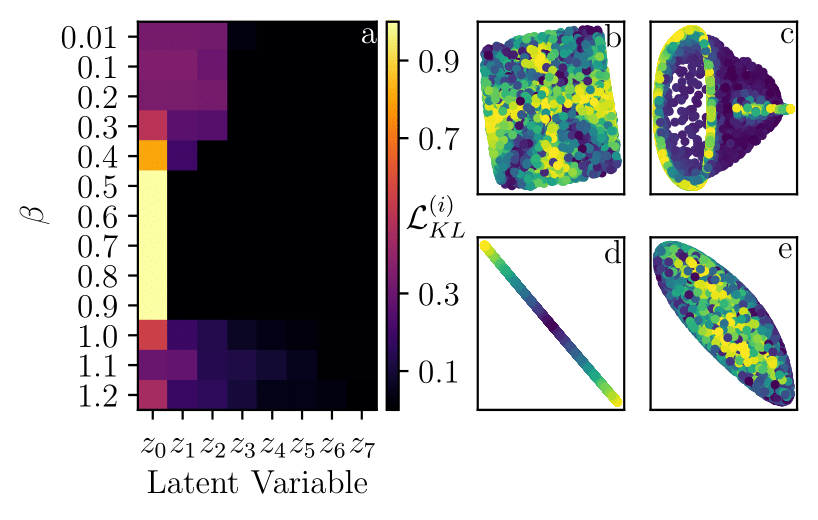}
    \caption{
    \textbf{Tuning the \boldmath$\beta$ parameter of \boldmath$\rho_s$-VAE leads to a compressed representation of quantum states.} a) Regularization loss $\mathcal{L}_{KL}^{(i)}$ contributed by each latent variable $z_i$ of $\rho_s$-VAE at different $\beta$ values. The $N=8$ latent variables are normalized and presented in descending order of loss values. b-e) Two-dimensional latent space $(z_0,z_1)$ of two largest $\mathcal{L}_{KL}^{(i)}$ at $\beta \in (0.01,0.4,0.75,1.0)$ values. The color-coding is identical to Fig. \ref{fig:exp_2} and indicates the concurrence value of the encoded quantum states.
    }
    \label{fig:exp_3}
\end{figure}

Fig. \ref{fig:exp_3}a visualizes the contribution of each latent variable to the regularization loss at different $\beta$ values. 
In this figure, each row is sorted and normalized. 
To interpret this visualization, we note that a regularization loss of $0$ corresponds to a latent variable $z_i$ that predicts the normal prior $\mathcal{N}(0,1)$ regardless of the input.
This is equivalent to not encoding information from the data.
Conversely, any deviation from $0$ regularization loss corresponds to a latent variable which encodes information.
We observe that at low regularization strengths $\beta \in [0.01,0.4]$, multiple latent variables contribute to the regularization loss.
In detail, Fig. \ref{fig:exp_3}b illustrates the two-dimensional latent space $(z_0, z_1)$ spanned by the two latent variables with the largest regularization losses $\mathcal{L}_{KL}^{(i)}$ at $\beta = 0.01$. 
Consistently with Fig.~\ref{fig:exp_2}b, both representations exhibit some observable structure according to the entanglement properties, but the information between the two axes is mixed.
As $\beta$ increases, the encoded information is increasingly concentrated in fewer latent variables.
This is because of the increased pressure on the latent variables to encode statistically independent features~\cite{burgessUnderstandingDisentanglingBeta2018}. 
In Fig.~\ref{fig:exp_3}c, for example, the two-dimensional latent space $(z_0, z_1)$ is shown at $\beta=0.4$, and we observe a clearer relationship between encoding and entanglement properties. 
In the critical region of $\beta \in[0.5,0.9]$, the number of active latent variables is equal to the number of ground-truth generative factors in the data set, namely one.
This means that majority of extracted information is represented in a single latent variable.
Fig. \ref{fig:exp_3}d shows the two-dimensional latent space $(z_0, z_1)$ at $\beta = 0.75$, where there is a direct relationship between encoding and entanglement properties.
Increasing $\beta$ above $0.9$ reduces the capacity of the latent variables to a point where the reconstruction quality becomes too poor to encode meaningful information. 
This leads to the latent variables becoming more similar to the prior again, as they encode a decreasing amount of information about the quantum states. 
This is visualized in \ref{fig:exp_3}e, where the two-dimensional latent space $(z_0, z_1)$ at $\beta = 1$ exhibits less observable structure again.

\subsection{Encoding Quantum States \boldmath$\rho_s(\alpha)$ with Tuned Regularization Strength \boldmath$\beta$} \label{chap:exp_4}
\begin{figure}[t]
    \centering
    \includegraphics[width=1.0\linewidth]{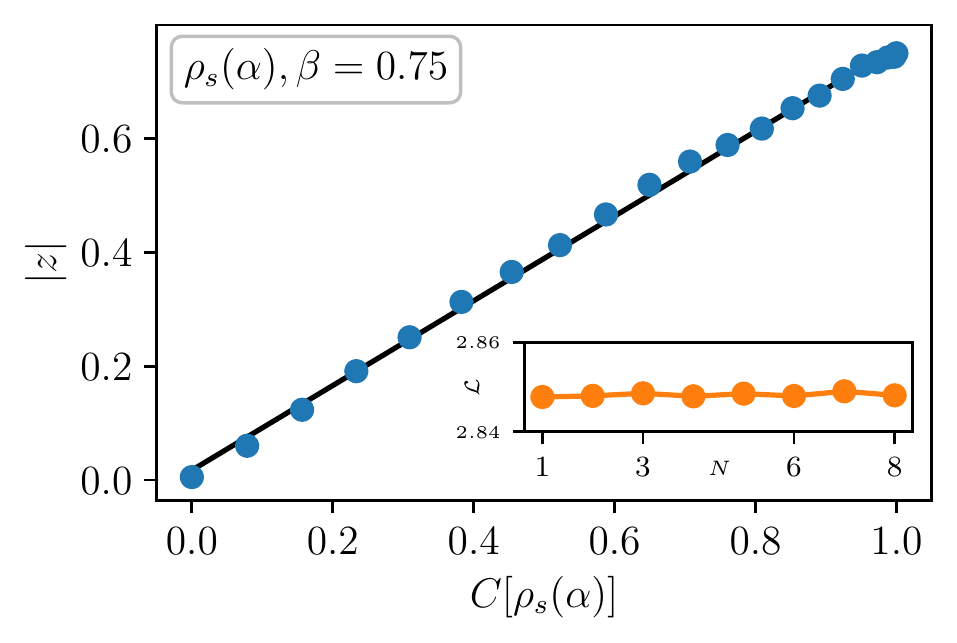}
    \caption{\textbf{The \boldmath$\rho_s$-VAE learns to extract concurrence from quantum states to structure its latent space.} \\  
    Correlation between one-dimensional latent space $|z|$ of $\rho_s$-VAE and concurrence $C[\rho_s(\alpha)]$ of encoded density matrices (blue, mean and standard deviation of $10$ samples). 
    The error bars are contained within the markers.
    The regression of encoded quantum states (black) shows a linear correlation with $r^2>0.99$.
    Inset: The final loss of $\rho_s$-VAE trained on quantum states $\rho_s(\alpha)$ at $\beta=0.75$ with latent space dimensions $N\in [1,8]$ (mean and standard deviation of $9$ experiments) indicates that a one-dimensional latent space has sufficient information capacity.  }
    \label{fig:exp_4}
\end{figure}

Based on the insight gained in the previous experiment, we proceed to analyze the $\rho_s$-VAE trained at $\beta=0.75$, where only a single latent variable $z_i$ is active.
Throughout the training process, we anneal the regularization strength from $\beta =0$ to $\beta =0.75$ to alleviate the problem of KL Vanishing~\cite{bowmanGeneratingSentencesContinuous2016a}.
The inset in Fig. \ref{fig:exp_4} shows that the one dimensional latent space is sufficient for this fixed $\beta$, which is what we expect from Fig.~\ref{fig:exp_3}a.  

The next step of the analysis is to examine and interpret the learned representation of the $N=1$ model to uncover what properties of quantum states are extracted to build the latent representation.
For this, we encode a test set of quantum states $\rho_s(\alpha)$ ($10$ samples at $21$ unique angles) using the trained $N=1$ model and record the predicted latent variable values $z$.

After encoding $\rho_s(\alpha)$ with the $N=1$ model, we plot the resulting latent variables against the concurrence $C[\rho_s(\alpha)]$ of the corresponding input in Fig.~\ref{fig:exp_4}.
As before, the resulting correlation is very close to linear, and we conclude that the learned mapping from input to latent representation is based on the extraction of entanglement information.
%
%
The understanding of why the model has learned to use this specific mapping from quantum state to latent representation starts with a comparison to the result in section \ref{chap:exp_1}. 
In this, the $\rho$-VAE with $\beta = 0$ and no information scrambling has learned to base its latent representation on the extraction of the angle $\alpha$, as this variable determines the underlying structure and enables the model to distinguish between the states.
By scrambling the density matrices, local properties such as the angle $\alpha$ become obscured. 
As a result, the $\rho_s$-VAE must extract a different quantity that contains equivalent information about the ground-truth generative factor to still be able to distinguish between quantum states.
Learning a function of $\alpha$ that remains invariant under the information scrambling transformation accomplishes this task, and the extraction of the concurrence $C[\rho_s(\alpha)] = C[\rho(\alpha)]$ does so.
A given angle $\alpha$ generates a unique concurrence $C[\rho_s(\alpha)]$ and thus provides a direct relation to the ground truth generative factor.

Finally, we remark that we report the absolute values $|z|$ rather than $z$, as the model has learned a representation with symmetry around $z=0$, which is a direct result of the regularization of latent variables. 
The unchanged latent space $z$ is presented in Appendix~\ref{chap:rhos}.
Since we are only interested in the relative distances of the quantum states in the encoding, this step does not remove the ability to interpret the latent representation.

\subsection{Testing the Ability of the \boldmath$\rho_s$-VAE to Generalize to Random Two-Qubit States}\label{chap:exp_5}
We proceed to explore the robustness and generalization capability of the representation learned by the $\rho_s$-VAE. Our objective is to determine whether the $\rho_s$-VAE can effectively extract entanglement information from any given pure (real) two-qubit state.

\begin{figure}[t]
    \centering
    \includegraphics[width=1.0\linewidth]{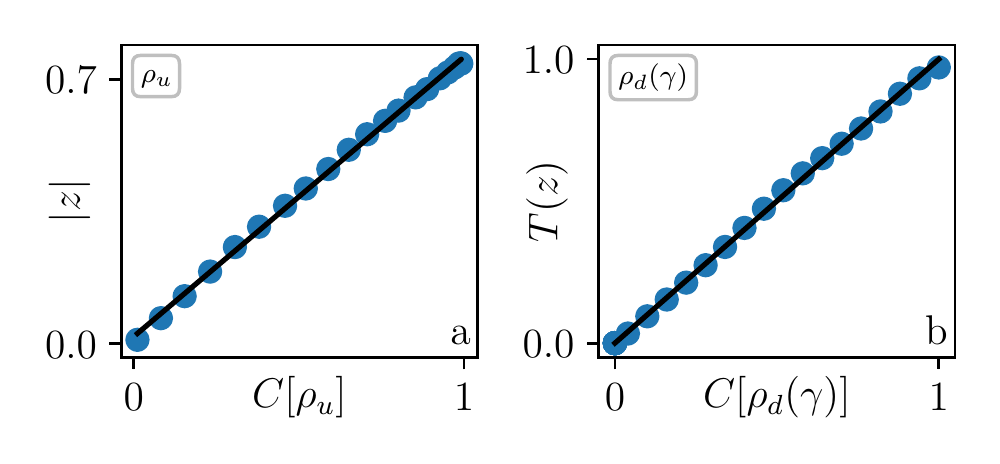}
    \caption{
    \textbf{The latent representation of the \boldmath$\rho_s$-VAE is able to generalize to other two-qubit systems.} \\ 
   Correlation between one-dimensional latent space $z$ of $\rho_s$-VAE and concurrence $C[\rho]$ of encoded density matrices (blue, mean and standard deviation of $10$ samples).
   The error bars are contained within the markers.
In this, the $\rho_s$-VAE is trained on $\rho_s(\alpha)$ and tested on a) states generated by random $4\times4$ unitaries $\rho_u$, and depolarized quantum states $\rho_d(\gamma)$. 
Both regressions of encoded quantum states (black) show that the correlation is linear with $r^2>0.99$.
    }
    \label{fig:exp_5}
\end{figure}

The information scrambling procedure in Eq.~\ref{eq:scramble} results in pure states that cover any real two-qubit state (see Appendix~\ref{chap:set_proof}), and hence the quantum states used for training and testing ($\rho_s(\alpha)$ for $\alpha \in [0,\pi]$ and $\rho_u$) belong to the same family of states. 
We therefore expect the model to work well on this task.
For this we test the model trained on $\rho_s(\alpha)$ with $N = 1$ and $\beta = 0.75$ on fully random two-qubit quantum states $\rho_u$ and record the predicted latent variable values $z$.
The set of density matrices $\rho_u$ comprises randomly generated two-qubit density matrices 
\begin{equation}
 \rho_u = U_{AB}\ket{00} \bra{00} U_{AB}^\dagger,
\end{equation}
where $U_{AB}$ represents the real components of randomly sampled $4 \times 4$ unitary operators, which are uniformly distributed according to the Haar measure.
Fig.~\ref{fig:exp_5}a illustrates the resulting correlation between mean predicted latent variable values (blue) and concurrence $C[\rho_u]$ of the corresponding input quantum states, showing that also for $\rho_u$ the encoding is linearly related to the concurrence.
In other words, the trained $\rho_s$-VAE is able to extract entanglement features from any pure (real) quantum state.

\subsection{Testing the Ability of the \boldmath$\rho_s$-VAE to Generalize to Depolarized Two-Qubit States}\label{chap:exp_6}
We now proceed to study mixed states $\rho_d(\gamma)$ obtained through a depolarization channel~\cite{nielsenQuantumComputationQuantum2010} starting from the maximally entangled state $\rho(\pi)$ 
\begin{equation}
    \rho_d(\gamma) = (1-\gamma)\rho(\pi) + \frac{\gamma}{2}1\!\!1. \label{eq:depol}
\end{equation}
In this transformation, $\rho(\pi)$ is mapped to a linear combination of the maximally mixed state and itself, and the degree of depolarization is set by $\gamma$.
A depolarization parameter of $\gamma=0$ produces a pure state and $\gamma=1$ produces the maximally mixed state.
We now encode quantum states $\rho_d(\gamma)$ for $\gamma \in [0, 1]$ using the trained $\rho_s$-VAE. 

Fig. \ref{fig:exp_5}b illustrates the correlation between (transformed) mean predicted latent variable values (blue) and concurrence $C[\rho_d(\gamma)]$ of the corresponding input quantum states at varying depolarization parameters $\gamma$.
In this, the transformation $T(z)$ re-scales the latent variable values 
\begin{equation*}
    T(z) = \max\left( 2.5 \frac{z - z_{\min}}{z_{\max} - z_{\min}}-0.5 ,0\right)
\end{equation*}
which is motivated in Appendix~\ref{chap:rho_d}.
We find that the model assigns latent variable values $T(z)$ that scale linearly with the concurrence, as demonstrated by the linear regression with $r^2>0.99$. 

This result is significant in that encoding the linear transformation of the maximally entangled state (Eq. \ref{eq:depol}) using the (highly) nonlinear $\rho_s$-VAE network leads to a latent representation that clearly shows the linear transformation of the input in an readily interpretable manner.
This observation that the $\rho_s$-VAE extracts a quantity that scales linearly with the depolarization process, in conjunction with the results of previous experiments, is compelling evidence that the $\rho_s$-VAE constructs its internal representation by extracting a quantity that is closely related to concurrence.

%



\subsection{Testing the Ability of the \boldmath$\rho_s$-VAE to Generalize to subsets of Three-Qubit States}
In the final step, we explore the capability of the trained $\rho_s$-VAE ($N=1$ and $\beta=0.75$) to investigate larger quantum systems. To achieve this, we examine quantum states $\rho_w(\alpha)$ generated by a parameterized three-qubit quantum circuit shown in Fig. \ref{fig:3q}. These states span a range from $\alpha=0$ (representing a separable state) to $\alpha = 2\arccos(\frac{1}{\sqrt{3}})$ (representing the W-state). We sample these states and record the corresponding two-qubit subpartitions $\rho_w^{AB}$, $\rho_w^{AC}$, and $\rho_w^{BC}$ for subsequent encoding using the $\rho_s$-VAE.

\begin{figure}[t]
    \centering
    \includegraphics[width=1.0\linewidth]{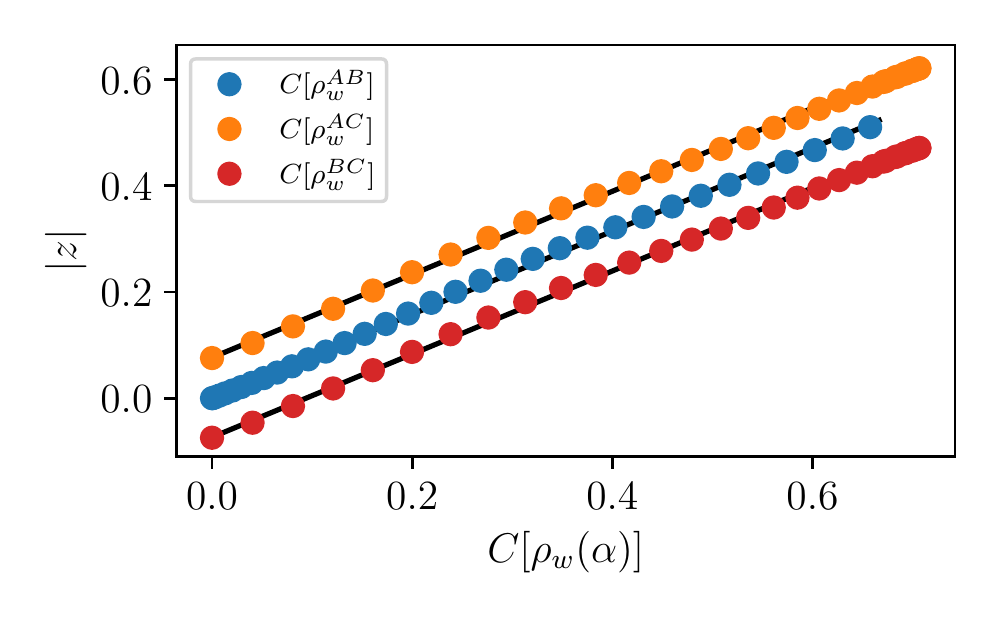}
    \caption{
    \textbf{The latent representation of the \boldmath$\rho_s$-VAE generalizes to subpartitions of three-qubit states.} \\ 
   Manual offset of correlation between one-dimensional latent space $|z|$ of $\rho_s$-VAE and concurrence $C[\rho_w(\alpha)]$ of encoded density matrices. 
In this, the $\rho_s$-VAE is trained on $\rho_s(\alpha)$ and tested on subparitions of the three-quit density matrices $\rho_w(\alpha)$. 
All regressions of encoded quantum states (black) show that the correlation is linear with $r^2>0.99$.
    }
    \label{fig:exp_6}
\end{figure}

Figure \ref{fig:exp_6} displays the correlation between the predicted latent variable values and the concurrence $C[\rho_w]$ for the three subpartitions. 
It is observed that the model assigns latent variable values that exhibit a linear scaling relationship ($r^2 > 0.99$) with the concurrence. This indicates that the model successfully generalizes to this system as well.

\section{Conclusion}\label{sec:discussion}
In this study, we investigate the use of the $\beta$-VAE framework for representation learning of small quantum systems. 
We focus on two-qubit density matrices generated by a parameterized quantum circuit, where the entanglement properties are determined by a single angle. 
By incorporating an information scrambling technique and optimizing the regularization strength, we observe that the VAE captures a quantity closely related to concurrence to structure its latent representation. 
Additionally, we demonstrate the generalization capability of the optimized model to other two-/three-qubit systems.
In conclusion, our findings establish the concept of employing machine learning techniques to derive interpretable representations for small quantum systems. 
These results serve as a solid foundation for future research endeavors, wherein the utilized methodology can be extended to investigate larger quantum systems.

\section*{Code availability}
The code used to train the model and perform
the analysis in this study has been made available at \href{https://github.com/FelixFrohnertDB/qcvae}{https://github.com/FelixFrohnertDB/qcvae}.

\section*{Data availability}
The data that support the findings of this study are openly available at \href{https://github.com/FelixFrohnertDB/qcvae}{https://github.com/FelixFrohnertDB/qcvae}.

\section*{Acknowledgments}
The authors would like to thank Vedran Dunjko, Mario Krenn, Jan Krzywda, Patrick Emonts, Adrián Pérez-Salinas, Simon Marshall, Stefano Polla, Tim Coopmans, Jordi Tura, and Aske Plaat for useful discussions.
This work was supported by the Dutch National Growth Fund (NGF), as part of the Quantum Delta NL programme.

\appendix

\section{Training Details} \label{chap:training_details}
We implement all models using Tensorflow's KERAS (2.11.0) machine learning package ~\cite{keras}
and train using the Adam optimizer ~\cite{kingmaAdamMethodStochastic2017a} for $1000$ epochs with a minibatch size of $64$. 
The initial learning rate is set to $5 \times 10^{-3}$ and is continuously reduced to a minimum of $10^{-4}$ with plateauing validation loss. 
Additionally, during training the $\beta$ value is slowly increased from $0$ to the final value  \cite{bowmanGeneratingSentencesContinuous2016a,fuCyclicalAnnealingSchedule2019}.
The training of all models was conducted on a CPU node within the Xmaris cluster, with each training session completed within a time frame of fewer than two hours.
The encoder and decoder architectures each consist of a fully connected MLP with $(16,8,4,2)$ hidden units in each respective layer and $\tanh$ as activation functions.
The encoder (decoder) network receives (produces) input (output) vectors consisting of $16$ entries, which represent a given density matrix.  
As a final detail, the models are trained on data sets comprising $101 \times 10^3$ quantum states.
For the generation of these training sets, we select $101$ angles within the range of $\alpha \in [0, \pi]$
and extract $10^3$ samples at each angle.


\section{Experiment Details}

\subsection{Quantum States \boldmath$\rho(\alpha)$} \label{app:calculation}
\begin{figure}[t]
    \centering
    \includegraphics[width=1\linewidth]{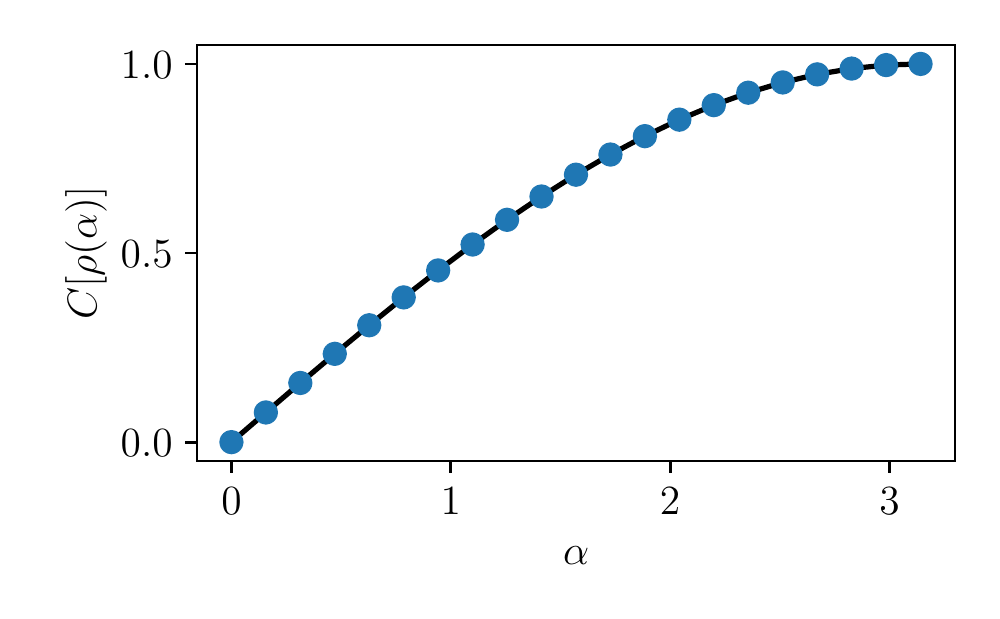}
    \caption{\textbf{The concurrence of quantum states \boldmath$\rho(\alpha)$.} \\
    Correlation between generative parameter $\alpha$ and concurrence of quantum states $\rho(\alpha)$ computed for $\alpha \in [0,\pi]$.
    }
    \label{fig:conc}
\end{figure}

The density matrix of output quantum states associated with the circuit in Fig. \ref{fig:overview} is the following:
\begin{equation}
    \rho(\alpha)  =\frac{1}{2}\begin{pmatrix}
    1 & \cos\frac{1}{2}\alpha &0& \sin\frac{1}{2}\alpha\\
    \cos\frac{1}{2}\alpha & \cos^2\frac{1}{2}\alpha & 0 & \sin\frac{1}{2}\alpha \cos\frac{1}{2}\alpha\\
    0&0&0&0\\
    \sin\frac{1}{2}\alpha& \sin\frac{1}{2}\alpha \cos\frac{1}{2}\alpha& 0&  \sin^2\frac{1}{2}\alpha 
    \end{pmatrix} \label{eq:dm}
\end{equation}
The concurrence $C[\rho(\alpha)]$, as defined in Eq. \ref{eq:conc}, at varying angles $\alpha \in [0,\pi]$ is presented in Fig.~\ref{fig:conc}.

\subsection{Quantum States with Information Scrambling \boldmath$\rho_s(\alpha)$ }\label{chap:rhos}
The information scrambling procedure introduced in section \ref{chap:exp_2} utilizes random unitaries to obscure local information from quantum states.
%
%
The algorithm consists of the following steps, as detailed in Lundberg et al. (2004) \cite{lundbergHaarMeasureGeneration2004a}.
 \begin{enumerate}
     \item Generate matrix $Z = X +iY$, where $X$, $Y$ are $2 \times
2$ matrices with entries that are normally
distributed with zero mean and unit variance
\item Compute QR-decomposition $Z = QR$
\item Compute diagonal matrix $\Lambda$ with $\Lambda_{i,i} = \frac{R_{i,i}}{|R_{i,i}|} $
\item Compute $U = Q\Lambda$ which is uniformly distributed
under Haar measure
 \end{enumerate}

\begin{figure}[t]
    \centering
    \includegraphics[width=1.0\linewidth]{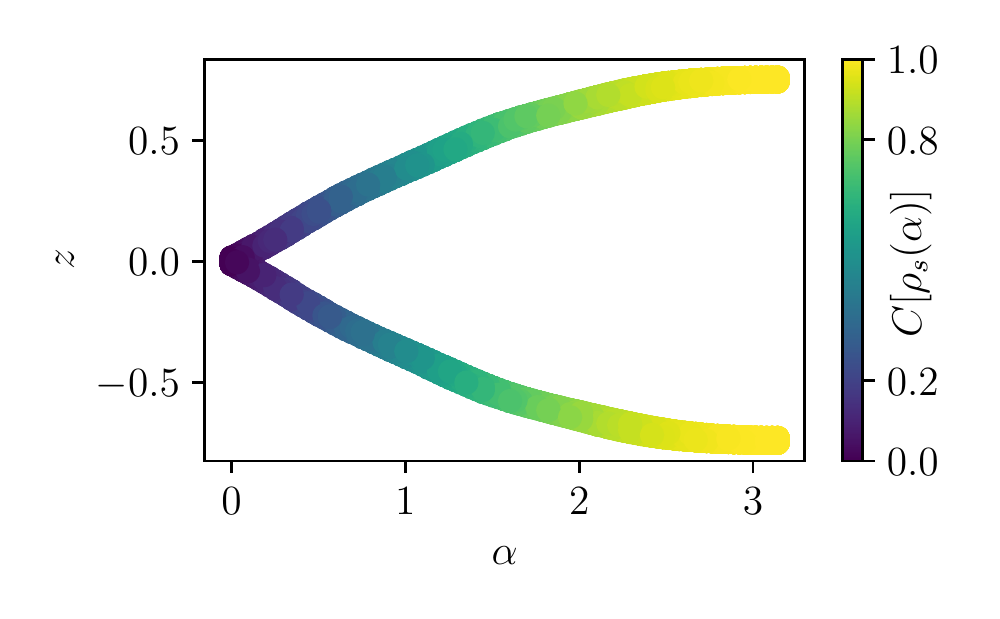}
    \caption{\textbf{The \boldmath$\rho_s$-VAE learns to extract concurrence from quantum states to structure its latent space.} \\  
    Correlation between one-dimensional latent space $z$ of $\rho_s$-VAE and generative parameter $\alpha$ of encoded density matrices. Each point is color-coded by its concurrence value.
      }
    \label{fig:rhos}
\end{figure}

In Fig. \ref{fig:exp_4}, we present an analysis of the predicted latent variables of the $\rho_s(\alpha)$ data set, focusing on their absolute values $|z|$. 
This presentation is necessary due to the symmetry around $z=0$ in the learned representation. 
For the sake of completeness, we include Fig. \ref{fig:rhos}, which illustrates the unchanged latent space $z$ as a function of $\alpha$.

\subsection{Random Unitary Quantum States \boldmath$\rho_u$}\label{chap:set_proof}
In section \ref{chap:exp_5}, we argue that the $\rho_s$-VAE is able to effectively generalize to the $\rho_u$ data set as its training data explores the whole pure (real) two-qubit state space.
To gain intuition for this statement, we first represent the density matrix $\rho_s(\alpha)$ in its state vector representation
\begin{align*}
    \ket{\psi_s(\alpha)} &= \frac{U_A \otimes U_B}{\sqrt{2}}  (\ket{00}+ 
    \cos(\frac{\alpha}{2})\ket{01}
    +\sin(\frac{\alpha}{2})\ket{11}) \\
\end{align*}
and apply the Schmidt decomposition
\begin{align*}
    \ket{\psi_s(\alpha)} &= \sum_i  \lambda_i(\alpha) U_A \ket{\psi^i_{1}}  \otimes U_B \ket{\psi^i_{2}}.
\end{align*}
The idea now is that with the rotation $U_A \ket{\psi^i_{1}}$ we can reach, by definition, any single qubit state $\ket{\phi^i_{1}}$.
By combining this with the ability to generate any entanglement value $\lambda_i(\alpha)$ with $\alpha \in [0,\pi]$ lets us explore the complete pure state space and can express any two-qubit state

\begin{equation*}
    \ket{\phi}=\sum \lambda_i' \ket{\phi^i_1}  \otimes \ket{\phi^i_2}.
 \end{equation*}
Hence, the underlying structure of quantum states used to train and test the $\rho_s$-VAE is identical, which makes the generalization possible.

\subsection{Quantum States with Depolarization \boldmath$\rho_d(\gamma)$}\label{chap:rho_d}
\begin{figure}[t]
    \centering
    \includegraphics[width=1.0\linewidth]{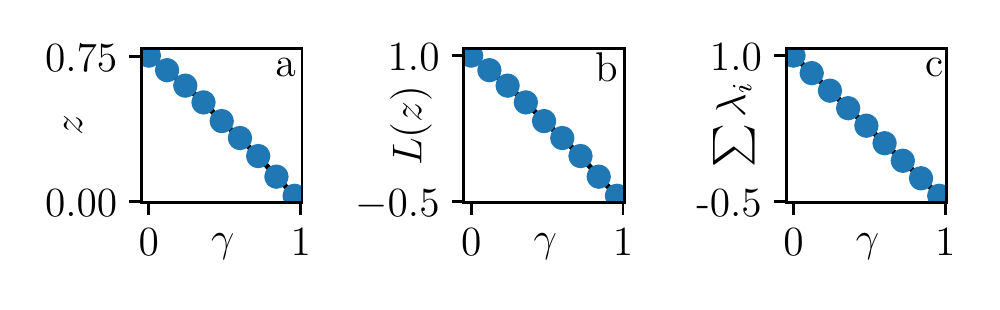}
    \caption{\textbf{The latent space of the \boldmath$\rho_s$-VAE generalizes to mixed states using the transformation $T(z)$. } \\  
    Correlation between the generative parameter $\gamma$ of encoded density matrices $\rho_d(\gamma)$ and
    a) the one-dimensional latent space $z$ of the $\rho_s$-VAE,
    b) the linear transformation $L(z)$ of the latent space of the $\rho_s$-VAE, and
    c) the sum of eigenvalues $\lambda_i$ of the Hermitian matrix $R$ in Eq. \ref{eq:r}.
    All regressions of the encoded quantum states (depicted in black) demonstrate a strong linear correlation with $r^2>0.99$.  }
    \label{fig:deco_un}
\end{figure}

In Section \ref{chap:exp_6}, we introduce the transformation $T(z)$ for the predicted latent variables to ensure the correct scaling with concurrence. For completeness, we include Figure \ref{fig:deco_un}a, which depicts the correlation between the unchanged latent variables and the depolarization parameter $\gamma$.
We observe that at $\gamma=0$, we encode the maximally entangled state $\rho(\pi)$ and obtain the same value for $z$ as shown in Fig. \ref{fig:exp_4}. 
As $\gamma$ increases linearly, there is a corresponding linear decrease in the predicted latent variables.
The point $\gamma = \frac{2}{3}$ marks the transition from entangled to separable quantum states, determined by the positive partial transpose (PPT) criterion~\cite{peresSeparabilityCriterionDensity1996}.
As discussed in section \ref{chap:exp_4}, the unchanged latent space accurately encodes the relative distances between encoded points in relation to the encoded density matrices, but the scaling is affected by the regularization of the latent encoding. 
To address this, we employ a linear transformation:
\begin{equation}
L(z) = 2.5 \frac{z - z_{\min}}{z_{\max} - z_{\min}}-0.5
\end{equation}
This transformation modifies the slope by a factor very close to $2$ and introduces offset to the latent variables ensuring that the maximally entangled state is encoded as $L(z)=1$ and the transition from entangled to separable occurs at $L(z)=0$. 
This is visualized in Fig. \ref{fig:deco_un}b.
Drawing inspiration from the definition of concurrence in Equation \ref{eq:conc}, we introduce the function $T(z) = \max(L(z),0)$ to achieve the desired performance.
The impact of excluding the $\max$ operation is illustrated in Fig. \ref{fig:deco_un}c, which exhibits identical results to Fig. \ref{fig:deco_un}b.

\subsection{Three-Qubit Quantum States \boldmath$\rho_w$}
\begin{figure}[t]
    \centering
    \includegraphics[width=1\linewidth]{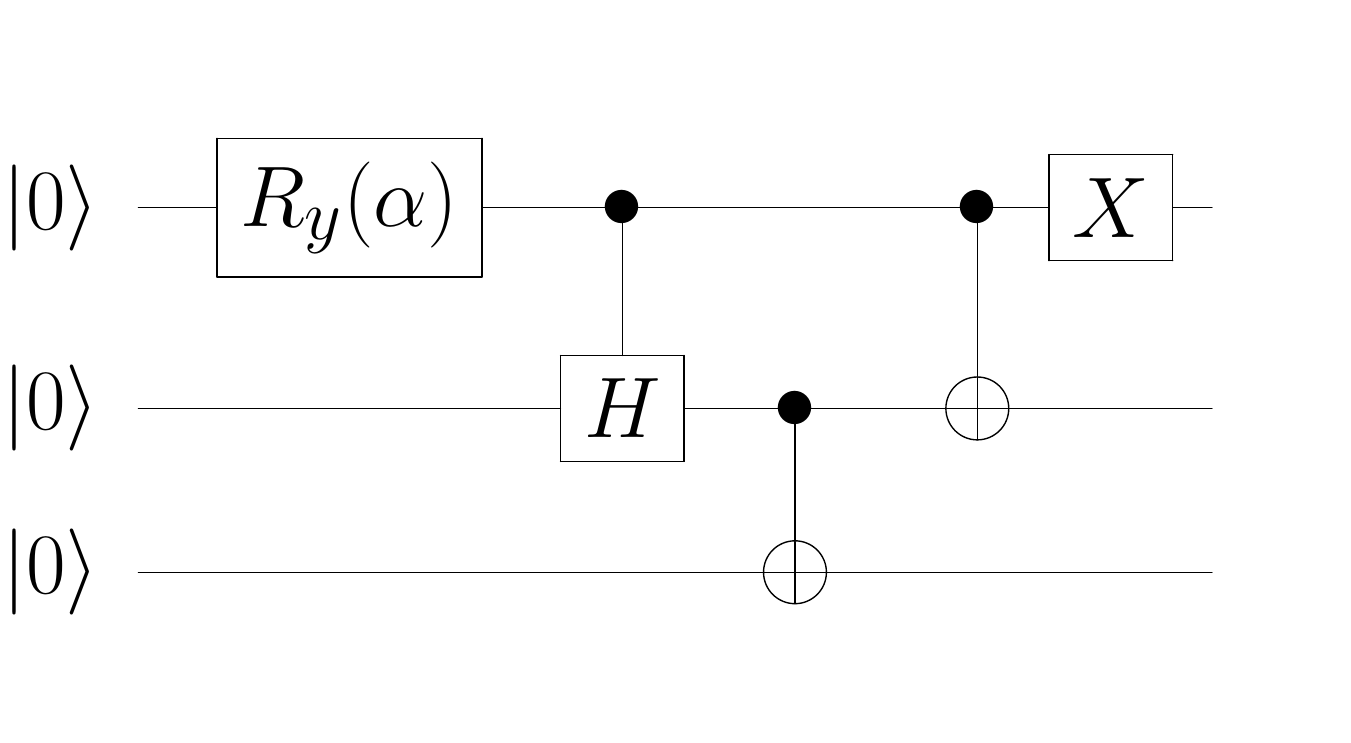}
    \caption{\textbf{3-qubit Parameterized Quantum Circuit.} \\
    Quantum states $\rho_w(\alpha)$ are generated by a three-qubit quantum circuit with a single parameterized $R_y(\alpha)$ gate.
    }
    \label{fig:3q}
\end{figure}

The $\rho_w(\alpha)$ data set utilized in Fig. \ref{fig:exp_6} consists of three-qubit states generated by the parametrized quantum circuit in Fig. \ref{fig:3q}.
This circuit is parameterized by a single parameter $\alpha$ which determines its entanglement properties:
For $\alpha=0$ the output state is separable and for $\alpha = 2\arccos(\frac{1}{\sqrt{3}})$ the output is the W-state.
To be able to use the model trained on two-qubit states, we subpartition the three-qubit states by performing a partial trace:
\begin{equation}
    Tr_A(\rho_w^{ABC}) = \rho_w^{BC}  
\end{equation}

\newpage

\bibliography{reference.bib}

\end{document}